# Increase of Complexity from Classical Greek to Latin Poetry

**R. Mansilla**               **E. Bush**

**Center for Interdisciplinary Research in Science and the Humanities**

**National Autonomous University of Mexico**

**Abstract**

In this paper we develop a method to analyze the increase of complexity from classical Greek poetry to classical Latin poetry by mapping large samples of those poetry onto a symbolic time series. This mapping setup intends characterize regular succession of rhythms, that is, the patterns of stressed and unstressed syllables in a verse. Using techniques from information theory, more precisely, certain Renyi entropy we show how the rhythmical patterns in Greek poetry evolve to more complex behavior in Latin poetry. Some interesting results are reported.

# 1 Introduction.

The human language is a system of objects, rules and relations of remarkable complexity. The purest expression of abstract and symbolic human imagery is without doubt poetry.

Without taking blank verse into account, poetry in English and Spanish is normally characterized by a regular succession of rhythm, that is, a pattern of stressed and unstressed syllables in a verse. Classical Greek and Latin poetry, on the other hand, not only has as a distinctive feature a chain of stressed and unstressed syllables, but also one of length, in other words, a pattern of "long" and "short" quantities of vowel length.

The study of symbolic sequences has a long history behind it. Gatlin's work [1] is probably the corner stone of this kind of studies. However, another pertinent publication which has exerted influence in the community is [2].

The discovery of the DNA structure has been influential in symbolic sequences studies, and has revolutionized our thoughts regarding biological evolution. The studies of long-range correlation in DNA [3-17] have been a fruitful line of research in recent years. A symbolic sequence is said to have long-range correlations if its power spectrum scales as $1/d^c$, where $d$ is the distance between symbols in the sequence and $c \approx 1$.

The success reached on account of the study of regularities of the DNA molecule has stimulated the study of written languages and music with similar techniques [18-22].

A recent paper in this direction is [23].

In the present paper, borrowing tools from information theory that have been used before in the DNA long-range correlation studies, we shall demonstrate that a particular type of rhythm in Greek and Latin poetry, the hexameter, can be analyzed with techniques similar to those reported above. We show that although both Greek and Latin poets used the hexameter the former did it in a more strict sense, while the Latin poets tended to "break the rules", creating (in reference to rhythm) more complex pieces of poetry. The evolving patterns of mutual and partial information functions, when applied to symbolic sequences mapped from real verses, allow us support our claims. The structure of the paper is as follows. In Sec. 2 we develop a brief overview of the ancient poets and their works analyzed here, as well as a description of the methodology used to map the verses into a symbolic sequence. In Sec. 3 we develop the main concepts and results related to information theory used in the paper. In Sec. 4 we discuss our results. Section 5 is the conclusion and Sec. 6 is for references.

## 2 An overview on ancient poetry.

We shall give a synoptic overview of the ancient poets analyzed here, as well as a description of the methodology used to map the verses into a symbolic sequence.

### 2.1 Classical Greek Poetry.

The hexameter, so named on account of it being a verse of six meters, or feet, was first used by Homer for the *Iliad* and the *Odyssey*[1], both epic poems that date back as far as the eighth century b.C. The former deals with the wrath of Achilles in the last year of the war against the Trojans, and the latter with the last year of the homecoming, travels and wanderings of Odysseus (Ulysses) ten years after the sack of Troy. Linguistically, the poem is packed with formulae, a string of words repeated in many places throughout both poems and with epithets, for example, "the grey-eyed Hera". Both poetical devices are used in order to meet the criteria of rhythm within a verse.

---

[1] There is a considerable amount of bibliography on whether Homer actually existed, and when and where the *Iliad* and the *Odyssey* were composed. For the purposes of the present paper, we have deliberately set the so-called "Homeric question" aside and have focused, instead, on considering the two extent poems as independent wholes. In a future paper we will discuss our result about this topic.

But the hexameter was not only limited to the singing of heroes and their deeds and travails in antiquity, it was also used in the genre of didactic and bucolic poetry. The first poet to use the hexameter for pedagogical and philosophical purposes was Hesiod, a poet whose worldview was radically different from Homer's. His *Works and Days* is didactic in nature given that his purpose in composing it was to teach his brother how to work a piece of land and which days were particularly propitious for agriculture. Hesiod's other poem, the *Theogony*, was an attempt at explaining the origins of the universe and the genealogy of the major and minor deities in ancient Greece. This systematization was also an attempt at giving the world, albeit in a mythical and religious context, a logical explanation.

Regarding the use of the hexameter for the genre of bucolic poetry, Theocritus is the model. His *Idylls* are characterized mainly by the description of bucolic situations in the countryside. Naturally, the urban element proper to the Hellenistic period of Greek history (323 b.C. - 30 b.C.), is absent from his poetry, with the exception of *Idyll* 15, and the emphasis is on the simple lifestyle of farmers. Theocritus is analyzed in this paper because Vergil, whom we will discuss in the section on classical Latin poetry, based some of his poems on the Greek poet from Syracuse.

It must be noted here that Homer and Hesiod, in their respective use of the hexameter, composed in the Ionian-epic dialect, an "artificial" dialect that was never actually spoken in the Greek-speaking areas of the Mediterranean, reserved only for poetry. It must also be remembered that using this dialect had religious and ritualistic purposes given that the epic was normally inspired by Calliope, the Muse of epic poetry. Theocritus, on the other hand, composed in the Doric dialect of ancient Greek, which was spoken in Syracuse. We should also remember that the nature of poetic language, in general, is to distance itself from everyday colloquial speech, hence the use of rhythm.

## 2.2 Latin Poetry.

With regards to classical Latin poetry, the poets analyzed here are Lucretius, Vergil and Ovid. We decided that these three poets represent the best models of the Latin hexameter, since they emulated the Greek models mentioned above.

Lucretius, whose work *On the Nature of Things*, set out to explain in hexameters the ordering of the universe and man's relation to it using the Epicurean philosophy as his base, especially the concept of atoms as the force that constitutes every type of being. His purpose was, like Hesiod's, didactic and philosophical, but what differentiates them is that the Roman approached the world from a more rational perspective, whereas Hesiod was more emotional.

Vergil was especially proficient in the use of the hexameter for his poetical works, the *Aeneid* and *Georgics.* The *Aeneid* relates the wanderings of Aeneas, son of Venus and a survivor of the destruction of Troy, and his quest to found a new homeland. It is considered by many philologists to have its model in the contents of the *Odyssey*, but it is also evident that Apollonious exercised also a good amount of influence on the Roman, since, like *Argonautica*, there is an absence of formulae and epithets in the Latin work. Vergil's objective was propagandistic, for the poem is filled with anachronistic references to the "greatness" of Rome. The *Georgics*, on the other hand, were didactic and, like Hesiod's *Works and Days*, dealt with descriptions of and advice on agriculture.

Ovid, Vergil's contemporary, was another Latin poet who found that the hexameter fit his poetical purposes. His best-known work is *Metamorphoses*, a compendium of myths of gods, semi-gods and heroes who were transformed, or suffered a metamorphosis, from one state of being to another. The poem is a wealth of information on mythology in the ancient world.

## 2.3 The hexameter.

As was noted at the beginning of this section, the hexameter in classical Greek and Latin poetry is rhythmical not only because of the regular succession of stressed and unstressed syllables, but also because each syllable can have a "long" or a "short" quantity. The rhythm of the hexameter is TA-ta-ta or TA-ta, depending on the combinations in each foot. But before we go any further, we shall look at the technical aspect of this type of verse.

The hexameter, as the etymology of the word indicates, is a verse of six meters, or feet, whose basic feature is the dactyl, that is, one long followed by two short syllables. Graphically, the verse looks like this:

$$\underline{\quad} \; uu \mid \underline{\quad} \; uu \mid \underline{\quad} \; uu \mid \underline{\quad} \; uu \mid \underline{\quad} \; uu \mid \underline{\quad} \; u$$

where $\underline{\quad}$ indicates a long syllable, and $u$ a short one. The dactyl $\underline{\quad} \; uu$ is thus represented. But there is another feature of the hexameter, the spondee, which is represented thus $\underline{\quad} \; \underline{\quad}$, that is, a long followed by another long. The vertical bar separates each foot from the other.

In the first four feet there can be either dactyls or spondees. The last foot also allows substitution of the short syllable for the long. The verse, then, is represented as follows with the possible substitutions:

$$\underline{\quad} \; \underline{uu} \mid \underline{\quad} \; \underline{uu} \mid \underline{\quad} \; \underline{uu} \mid \underline{\quad} \; \underline{uu} \mid \underline{\quad} \; uu \mid \underline{\quad} \; u$$

Notice that the fifth foot does not allow any substitution. This, generally speaking, occurs in the Greek and Latin poets. However, when the dactyl of the fifth foot is substituted with a spondee, the fourth foot without exception is a dactyl. Each foot has a princeps and a biceps, the former being their first half, always long, and the biceps being the second half, either two shorts or one long. The princeps is, rhythmically speaking, characterized by the arsis, or the up beat, and the biceps by the thesis, or the down beat, resulting in the rhythm, as was seen above, of TA-ta-ta or TA-ta.

Another essential feature of the hexameter is the caesura, or pause. There is widespread disagreement on how the caesura is really used: as a rhythmical pause, allowing a poet to, as it were, "catch his breath", or as a semantic marker, that is, a device used to call attention to a certain word or turn of phrase within the verse. Either way, Greek and Latin poets followed certain rules regarding the placement of the caesura.

Normally, the caesura is placed at a word-end within the foot, and it can also be placed so as to coincide with a word-end and a foot-end. The placement depends on the foot, that is, the position is normally after the princeps or between the biceps when there is no substitution. In the first case, the caesura is "masculine", and in the second it is "feminine". The most common caesurae are known as trimimeral, penthemimieral and hephthemimieral, respectively, which is the same as saying they fall after the third, fifth and seventh princeps. There is also the bucolic caesura, which coincides with the end of the fourth foot. An example of an analysis of a verse, also called scansion, is taken from the first verse of Book I of the Aeneid:

*Arma uirumque cano, Troiae qui primus ab oris*

and it is analyzed thus

$$\underline{\phantom{m}} \; uu \; | \; \underline{\phantom{m}} \; uu \; | \; \underline{\phantom{m}} \; \Uparrow \; \underline{\phantom{m}} \; | \; \underline{\phantom{m}} \; \underline{uu} \; | \; \underline{\phantom{m}} \; uu \; | \; \underline{\phantom{m}} \; u$$

The double-bar arrow indicates a caesura, which in this case, is penthemimeral. It is placed at a word-end, *cano*, and coincides with the comma. The rhythm of the verse is read as follows:

*ARma uirUMque caNO, TroiAE qui PRImus ab Oris* (1)

or

TA-ta-ta TA-ta-ta TA ⇑ ta TA-ta-ta TA-ta-ta TA-ta (2)

Generally, if the penthemimeral appears, there are no other caesurae. If there is a trimimeral, it is usually corresponded by the hephthemimeral or the bucolic caesura. These rules are not airtight, since there are instances in which punctuation (commas, periods etc.) plays an important role and can be found in the first foot, thus indicating a caesura.

The possible combinations within a foot are

__ u u , __ __ , __ ⇑ u u , __ u ⇑ u , __ ⇑ __ , __ u u ⇑ , __ __ ⇑ , __ u.

In this paper, all the elements of the hexameter are mapped into symbolic time series using a trinary system where a long syllable is translated as 0, a short as 1 and the caesura as 2. The verse (1)-(2) analyzed above is thus 0110110200101101. In order to have representative samples of each poet, we decided that the first one-hundred verses of each Greek and Latin poet were sufficient. However, two things should be noted: first, we did not take the grammatical and syntactic aspects into account; second, further study is forthcoming in which the complete works of the Greek and Latin poets mentioned here will be analyzed so as to arrive at more definite and sounder conclusions (see footnote in page 3).

At the risk of sounding repetitive, the poets and their works analyzed here are: Homer, the first hundred verses of Books I of the *Iliad* and the *Odyssey*; Hesiod, the first hundred verses of *Works and Days*; Theocritus, the first hundred verses of Book I of *Idylls*. The Latin poets and their works are: Vergil, the first hundred verses of Book I of the *Aeneid* and the first hundred verses of Book I of the *Georgic*; Ovid, the first hundred verses of Book I of *Metamorphoses*; Lucretius, the first hundred verses of Book I of *On the Nature of Things*. The texts used are from the Oxford editions.

## 3 The mutual information function.

The concept of mutual information function can be found for the first time in the epic 1948 paper by Shannon [24]. His results were generalized to abstract alphabets by several Soviet mathematicians, culminating in the work of Dobrushin [25]. Initially they were used to measure the difference between the average uncertainty in the input of an information channel before and after the output is received. More recently the method has been used to study some properties of regular languages and cellular automata [26], DNA long-range correlation [1], [6-8], [15-16] and some properties of strange attractors [27]. A thorough discussion of the properties of this function can be seen in [28-30].

Let us denote by $A = \{a_1, \ldots, a_n\}$ an alphabet and $s = (\ldots, a_0, a_1, \ldots)$ an infinite string with $a_i \in A, i \in Z$, where $Z$ represents the set of all integer numbers. The mutual information function of the string $s$ is defined as:

$$M(d,s) = \sum_{a,b \in A} P_{a,b}(d,s) \ln \left[ \frac{P_{a,b}(d,s)}{P_a(s) P_b(s)} \right] \tag{3}$$

where:

$P_{a,b}(d,s)$ is the joint probability of having the symbol $a$ followed $d$ sites away by the symbol $b$ on the string $s$.

$P_a(s)$ is the density of the symbol $a$ in the string $s$.

We will also call partial $(a, b)$ information function to:

$$M_{a,b}^P(d,s) = P_{a,b}(d,s) \ln \left[ \frac{P_{a,b}(d,s)}{P_a(s) P_b(s)} \right] \tag{4}$$

Notice that function defined in (3) represents the sum of the contribution of functions defined in (4) for every pair $(\boldsymbol{a}, \boldsymbol{b}) \in A^2$. When no confusion is possible we drop the symbol $s$ from (3) and (4), resulting the following equations:

$$M(d) = \sum_{\boldsymbol{a}, \boldsymbol{b} \in A} P_{\boldsymbol{a}, \boldsymbol{b}}(d) \ln\left[\frac{P_{\boldsymbol{a}, \boldsymbol{b}}(d)}{P_{\boldsymbol{a}} P_{\boldsymbol{b}}}\right] \qquad (5)$$

$$M_{\boldsymbol{a}, \boldsymbol{b}}^{P}(d) = P_{\boldsymbol{a}, \boldsymbol{b}}(d) \ln\left[\frac{P_{\boldsymbol{a}, \boldsymbol{b}}(d)}{P_{\boldsymbol{a}} P_{\boldsymbol{b}}}\right] \qquad (6)$$

In our work we consider the alphabet $A = \{0, 1, 2\}$ and obviously our sequences are not infinite, but large enough to allow stable statistical estimations of $P_{\boldsymbol{a}, \boldsymbol{b}}(d)$ and $P_{\boldsymbol{a}}$.

As we will show below functions (5) and (6) often have periodic behavior. Therefore, we briefly describe here a method widely used to study those kinds of behavior. Fourier spectra [31] is used in time series analysis, because the visual representation in the frequency domain can more easily reveal patterns that are harder to discern in the primary data, for example, intricate periodical behavior. We use here Fourier transform of mutual and partial information functions to detect some periodical behavior of those functions when applied to the symbolic strings mapped from the verses. From this point forward, we call power spectra of mutual (partial) information function to the product of Fourier transform of those functions by its complex conjugate:

$$\hat{S}(k) = \boldsymbol{q} \left| \sum_{d=1}^{L} M(d) e^{-2 \boldsymbol{p} i (k/L) d} \right|^2$$

where $\boldsymbol{q}$ is a constant related with the sample frequency and $L$ is the number of data available for $M(d)$.

**4 Discussion.**

We begin our analysis with the behavior of mutual information function related to each verse. It drastically changes from Greeks to Latin poets. In Fig. 1a the mutual information function of Homer's *Iliad* and in Fig. 1b that of Vergil's *Aeneid* is shown. They represent the average behavior for Greek and Latin poems respectively. The peak at distance $d = 2$ is more pronounced in Vergil than in Homer. It is related to the long syllable common to the dactyl and the spondee and the substitution between them. The peak at distance $d = 9$ in Homer's *Iliad* almost disappears in Vergil's *Aeneid*. Therefore, some structures inside the verses were relaxed in Latin poetry with reference to the Greek one. Below, we study these facts more accurately. Recall that mutual information functions (5) is the sum of the partial contributions (6). Finally, the peak at distance $d = 17$ in Homer's *Iliad*, moves to $d = 16$ in Vergil's *Aeneid* and is more pronounced here. As we show below, this fact is related to the use and position of caesurae.

We will now take a closer look at the structure of the verses using partial information functions. A remarkable property of functions $M_{0,1}^{P}(d)$ and $M_{1,1}^{P}(d)$ is that their graphs are a mirror image of each other with respect to the horizontal axis. In Fig. 2a the $M_{0,1}^{P}(d)$ and $M_{1,1}^{P}(d)$ functions of Homer's *Iliad* are shown. In Fig. 2b and Fig. 2c the $M_{0,1}^{P}(d)$ and $M_{1,1}^{P}(d)$ functions for the Greek and Latin poets, respectively, are shown. An important consequence of this fact is that $M_{0,1}^{P}(d) + M_{1,1}^{P}(d) = 0$ and hence they contribute nothing to the mutual information function. Therefore peaks at distances $d = 2,9$ in mutual information functions of every poem are contribution of the remainder partial information functions.

The relationship between the princeps of the feet and caesurae is also different in Greek and Latin poetry. In Fig. 3a and 3b the $M_{0,2}^P(d)$ functions for Greek and Latin poems, respectively, are shown. Notice the peak at distance $d = 9$ in both set of plots. The height is larger in Greek poems than in Latin ones. The peak at distance $d = 13$ almost disappears in Latin poems. These facts suggest that the use of more than one caesura in each verse is more often in Latin than in Greek poems. Besides, the rhythmical structure in Greek poems encompasses several verses as the peaks at distances $d = 26, 44$ show, which almost disappears in Latin poems.

The distance between two consecutive caesurae also yields some insights about how more complex Latin poetry is. In Fig. 4a,b,c $M_{2,2}^P(d)$ for Homer's *Iliad*, Homer's *Odyssey* and Lucretius' *Nature of Things* is shown. Notice the remarkable periodic behavior of $M_{2,2}^P(d)$ for *Iliad* that maintains high correlation between caesurae at distance of 170 symbols. This feature disappears in Lucretius' verse. The periodic structure persists however in Latin poets, although it is weakened. In Fig. 5 the averaged power spectrum of Greek and Latin poems is shown. The first large peak almost coincides in position, but is smaller in Latin poem, reflecting a lower influence of the corresponding harmonic. Our explanation of this fact is that rhapsodes in Homeric times had to memorize large pieces of verses, because no written verses existed at the time. Therefore, a strict rhythm facilitated memorization. Meanwhile, in Lucretius' time declaimers also memorized verses, but composition could be done by writing them. It allows more complicated structures in the poetry.

With reference to the *Odyssey*, it proved to be very flashy how the same person could write two epic poems in the same period with that striking difference in rhythm. The behavior of *Odyssey* is closer to Lucretius' Nature of Things than to Homer's *Iliad*. Our results probably could throw light over the question if *Iliad* and *Odyssey* are of the same author.

## 5 Conclusions.

In our work we have shown that an increase in the complexity of rhythmical structure arises in the use of hexameter from Greek classic poets to Latin ones. Our method allows the discrimination between Greek and Latin poems using the differences in the use of dactyl and spondee and in the position and use of caesurae. Our results also open the possibility of deciding on the Homeric question.

# 6 References.

[1] Gatlin, L. L.; "Information Theory and the Living System", Columbia University Press (1972).

[2] Herzel, H.; "Complexity of symbol sequences", Sys. Anal. Mod. Sim., **5** (1988), 435-444.

[3] Borsnitk, D, *et al.*; "Analysis of apparent $1/f^a$ spectrum in DNA sequences", Europhysics Letters, **23** (1993) 389-394.

[4] Buldyrev, A. L., *et al.*, "Long-range correlation properties of coding and noncoding DNA sequences: GenBank analysis", Physical Review E **51** (1995) 5084-5091.

[5] Karlin, S., Brendel, V.; "Patchiness and correlations in DNA sequences", Science, **259** (1993) 677-680.

[6] Li, W.; "Generating nontrivial long-range correlation and $1/f$ spectra by replication and mutation", International Journal of Bifurcation and Chaos, **2** (1992) 137-154.

[7] Li, W., Kaneko, K.; "Long-range correlation and partial $1/f^a$ spectrum in a noncoding DNA sequence", Europhysics Letters, **17** (1992) 655-660.

[8] Miramontes, P; "Cellular Automatas, Genetic Algorithms and DNA evolution", Ph D. thesis UNAM, 1992.

[9] Peng, C. K., *et al.*; "Long-range correlation in nucleotide sequences", Nature, **356** (1992) 168-170.

[10] Nee, S.; "Uncorrelated DNA walks", Nature, **357** (1992) 450.

[11] Prabhu, V. V., Claverie, J. M.; "Correlations in intronless DNA", Nature, **359** (1992) 782.

[12] Li, W., Kaneko, K.; "DNA correlations", Nature, **360** (1992) 635-636.

[13] Chatzidimitrou-Dreismann, C. A., Larhamar, D.; "Long-range correlation in DNA", Nature, **361** (1993) 212-213.

[14] Peng, C. K., *et al.*; "Finite size effects on long-range correlations: implications for analyzing DNA sequences", Physical Review E, **47** (1993) 3730-3733.

[15] Mansilla, R, Mateo- Reig, R.; "On mathematical modeling of intronic sectors of the DNA molecule", International Journal of Bifurcation and Chaos, **5** (1995) 1235-1241.

[16] Voss, R.; "Evolution of long-range correlations and $1/f$ noise in DNA base sequences", Physical Review Letters, **68** (1992) 3805-3808.

[17] Mansilla, R, Cocho, G.; "Multiscaling in expansion modification systems: an explanation for long range-correlation in DNA", Complex Systems, **12** (2000) 207-240.

[18] Voss, R., Clarke, J.; "1/f noise in music and speech", Nature, **258**, 317-318 (1975).

[19] Klimontovich, Yu., Boon, J.; "Natural flicker noise (1/f noise) in music", Europhysics Letters, **3**, (1987) 395-399.

[20] Schenkel, A., Zhang, J., Cheng, Y.Ch.; Fractals, **1**, (1993) 47-52.

[21] Shi, Y.; "Correlations of pitches in music", Fractal, **4**, (1996) 547-553.

[22] Saiki, T. *et al.*; "Fluctuations of character centroid intervals in laterally written Japanese sentences", IEICE Transactions on Fundamentals of Electronics, Communications and Computer Sciences, E82A(3) (1999) 520-526.

[23] Montemurro, M. A., Pury, P. A., "Long-range fractal correlations in literary corpora", cond-mat/0201139 (to appear in Fractals).

[24] Shannon, C.; "A mathematical theory of communication", Bell Systems Technologies Journal **27**, (1948) 379-423.

[25] Dobrushin, R.; "General formulation of Shannon's main theorem in Information Theory", Uspieji Matematicheskix Nauk **14**, (1959) 1-104. English translation in American Mathematical Society Transaction **33**, (1959) 323-438.

[26] Li, W.; "Power spectra of regular languages and cellular automata", Complex Systems **1**, (1987) 107-113.

[27] Fraser, A., Swinney, H.; "Independent coordinates for strange attractors from mutual information", Phys. Rev. A **33**, (1986) 1134-1151.

[28] Blahut, R.; "Principles and practice of information theory", Addinson-Wesley, (1987).

[29] Li, W.; "Mutual information function versus correlation function", Journal of Statistical Physics **60**, (1990) 823-831.

[30] Renyi, A.; "The probability theory", Pergamon Press, (1970).

[31] Percival, D. B., Walden, A. T.; Spectral Analysis for Physics Applications, Cambridge University Press, 1993.

**Figures captions.**

Fig. 1: Mutual information functions of Homer's *Iliad* (a) and Vergil's *Aeneids* (b).

Fig. 2: Partial information functions $M_{0,1}^P(d)$ and $M_{1,1}^P(d)$ for Homer's *Iliad* (a), all the Greek poems (b) and all the Latin poems (c). Notice the mirror behavior of the partial information functions (see the text).

Fig. 3: Partial information functions $M_{0,2}^P(d)$ for Greek poems (a) and Latin poems (b).

Fig. 4: Partial information functions $M_{2,2}^P(d)$ for Homer's *Iliad* (a), Homer's *Odyssey* (b) and Lucretius' *Nature of Things* (c).

Fig. 5: Averaged power spectrum for Greek and Latin poems.

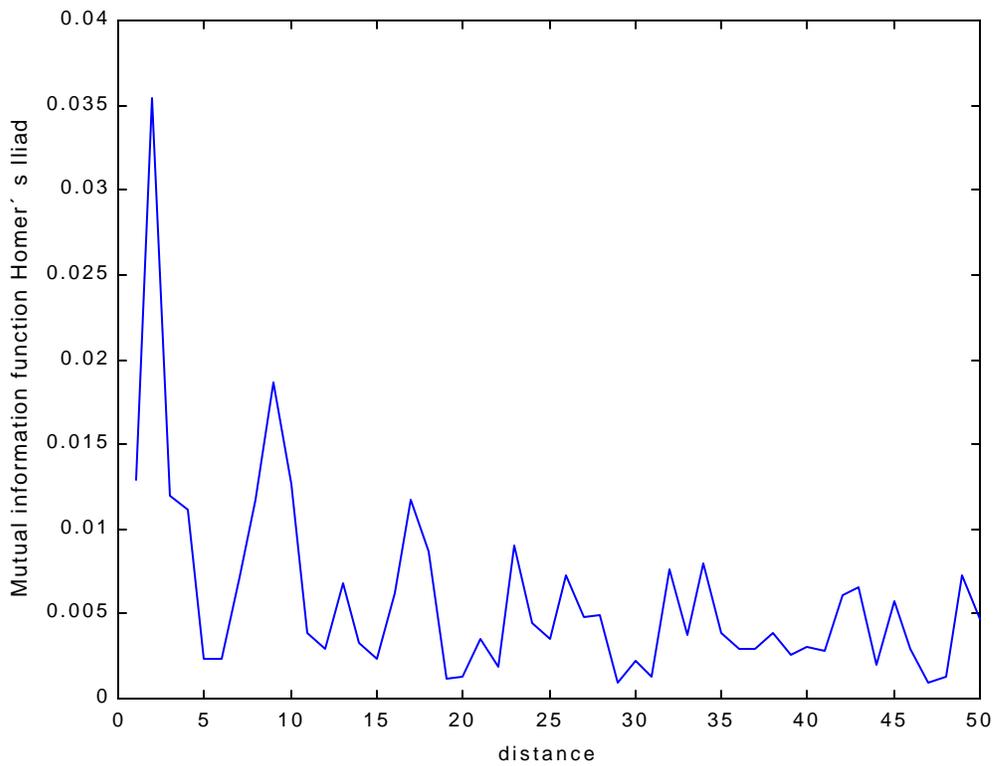

**Fig. 1a Mansilla & Bush**

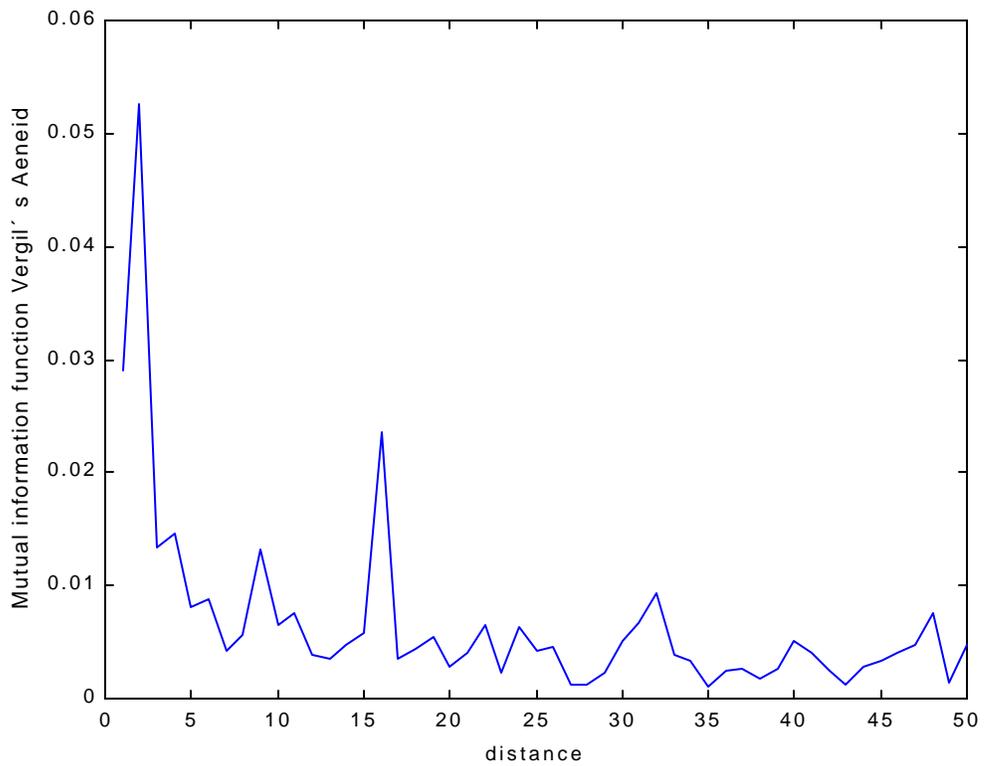

**Fig. 1b of Mansilla & Bush**

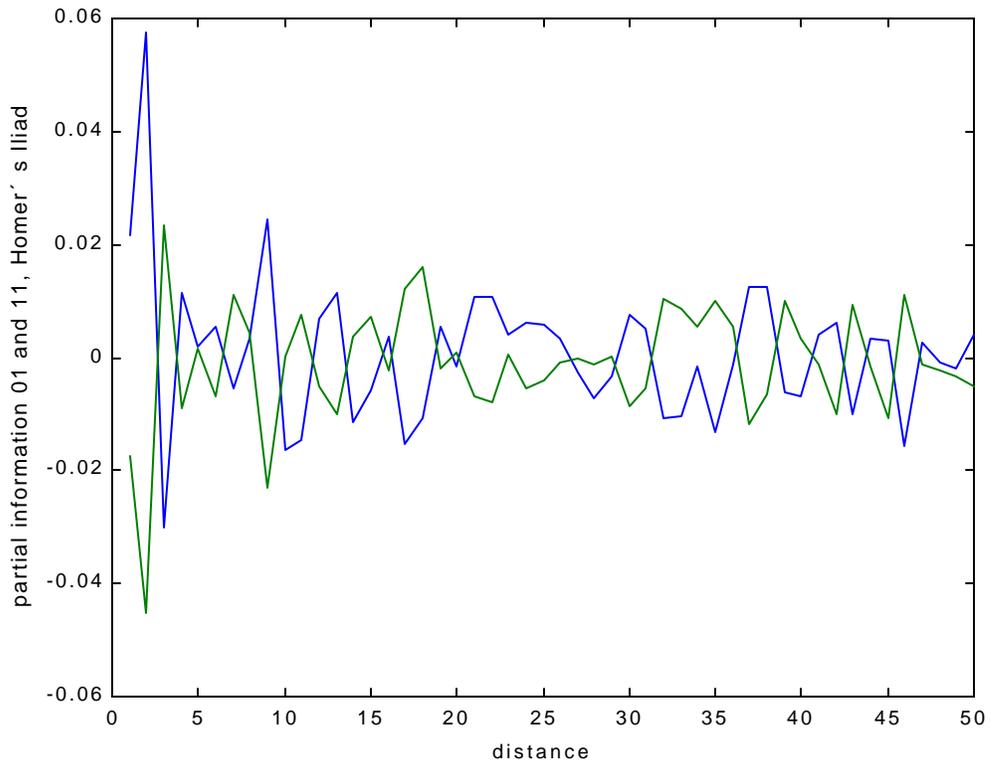

**Fig. 2a Mansilla & Bush**

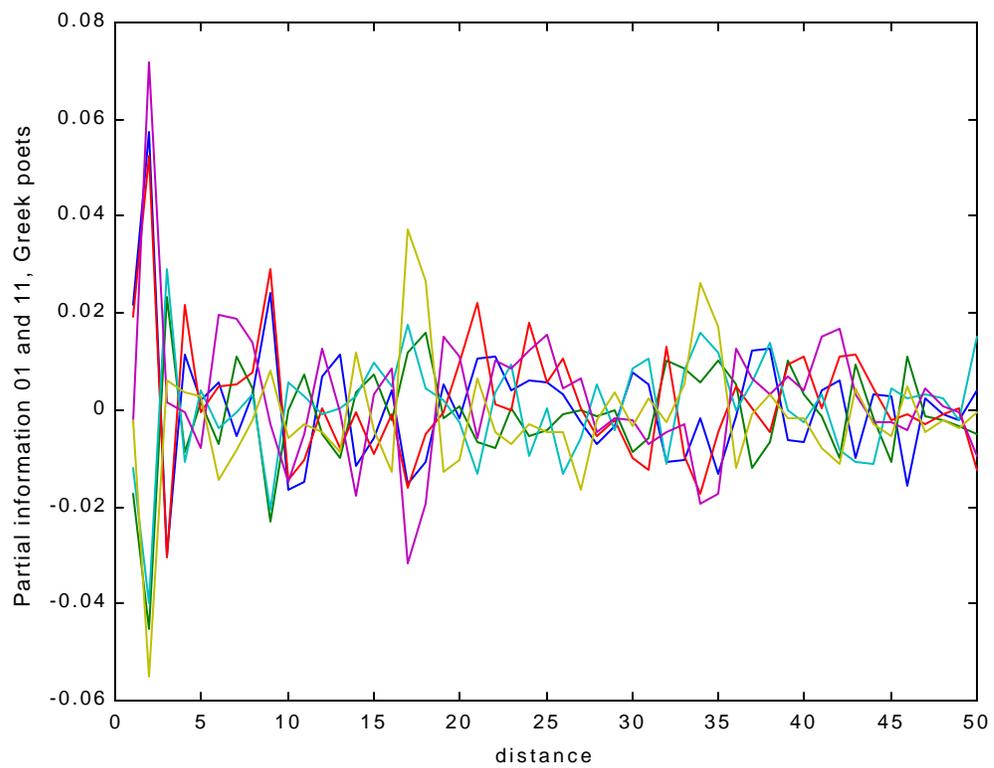

**Fig. 2b Mansilla & Bush**

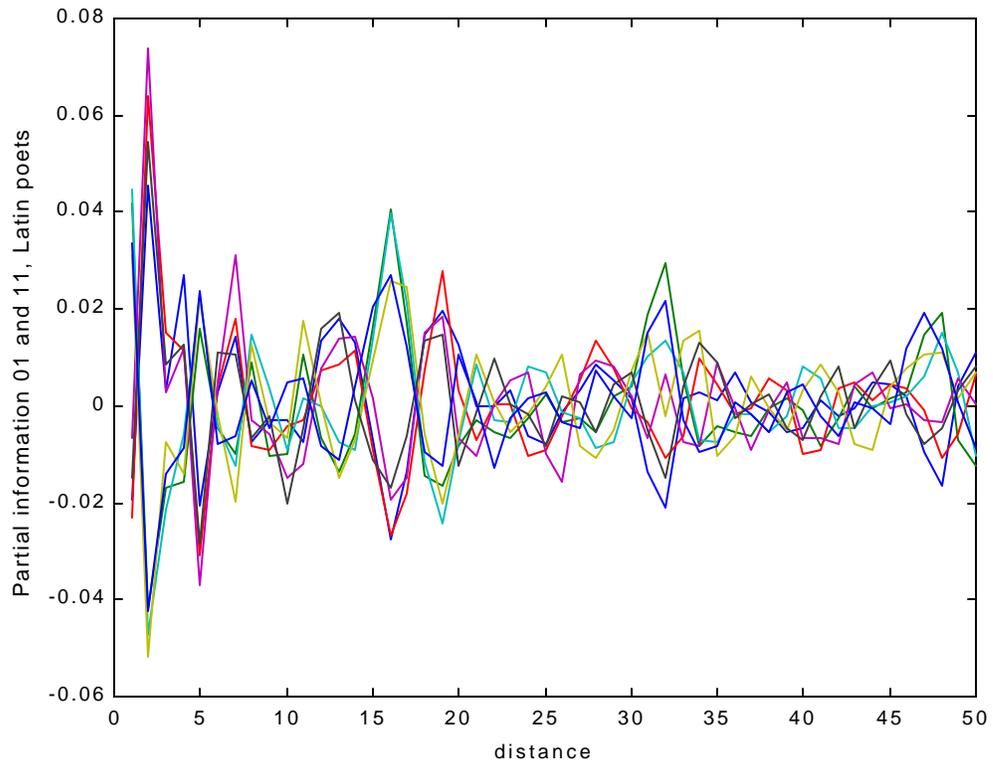

**Fig. 2c Mansilla & Bush**

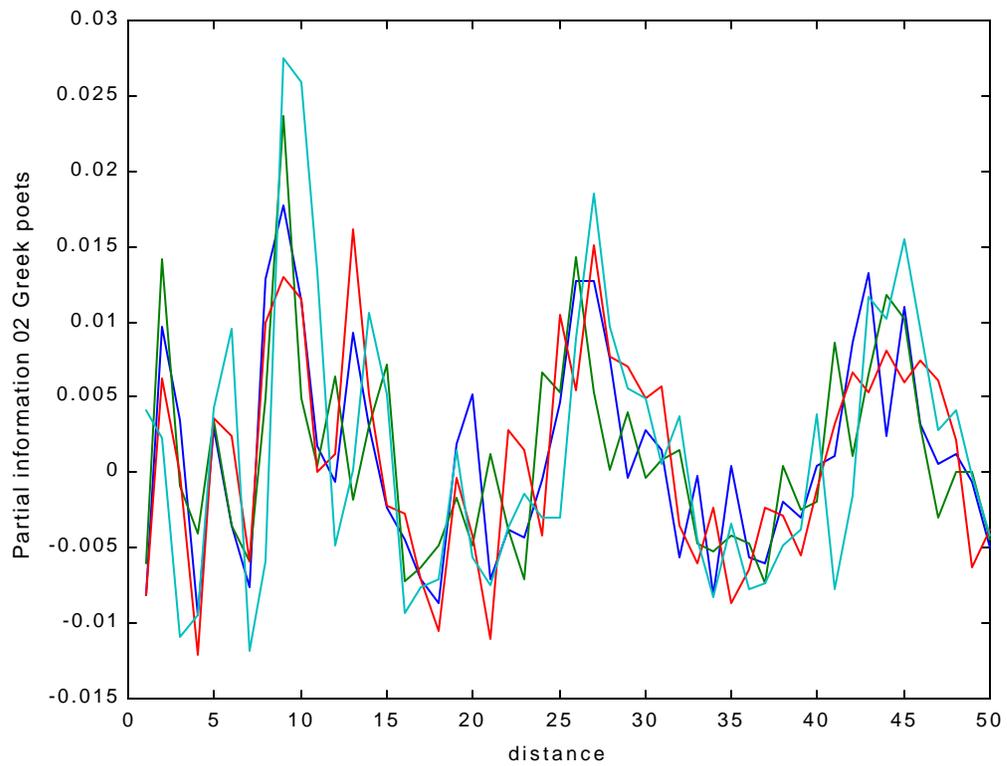

**Fig. 3a Mansilla & Bush**

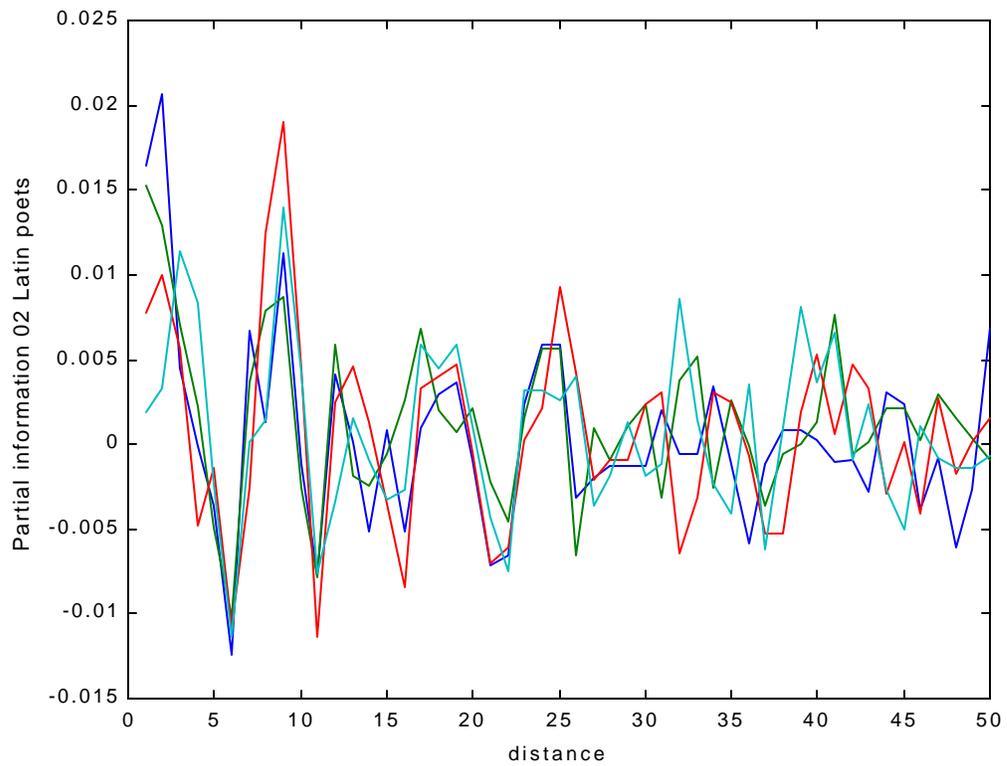

**Fig. 3b Mansilla & Bush**

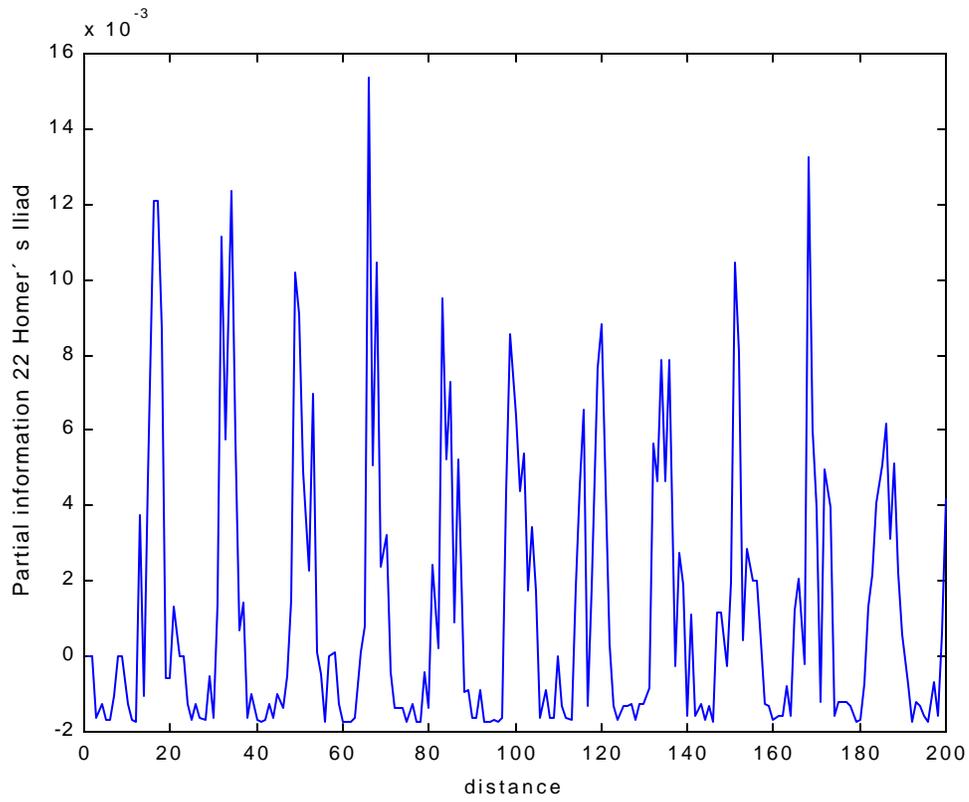

**Fig. 4a Mansilla & Bush**

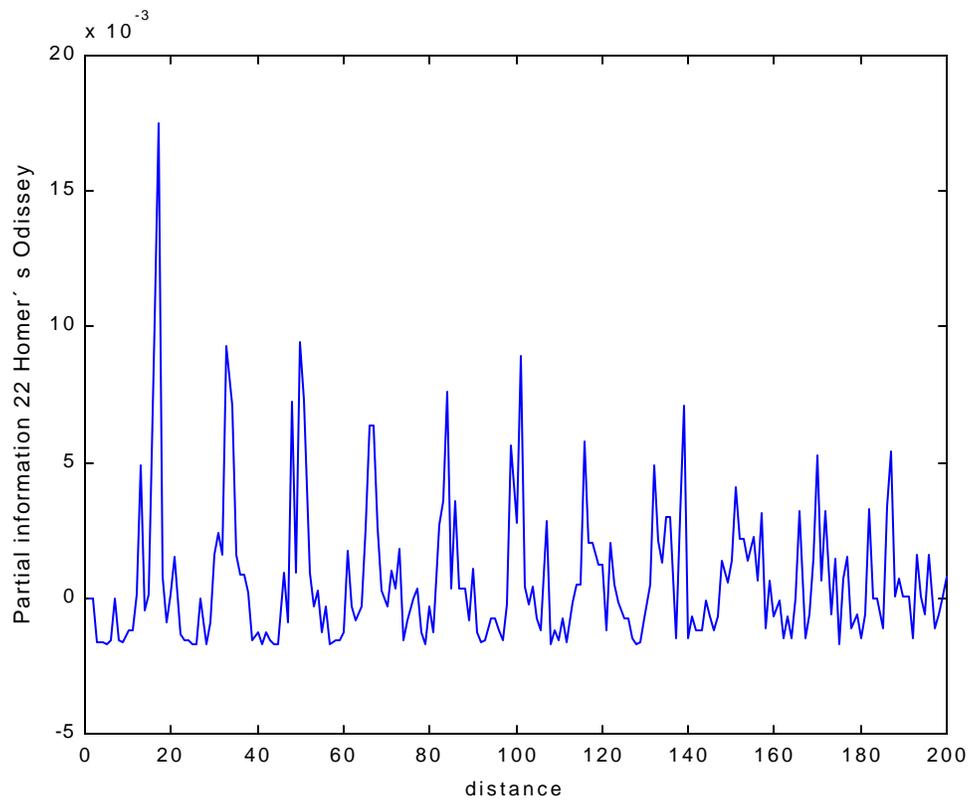

**Fig. 4b Mansilla & Bush**

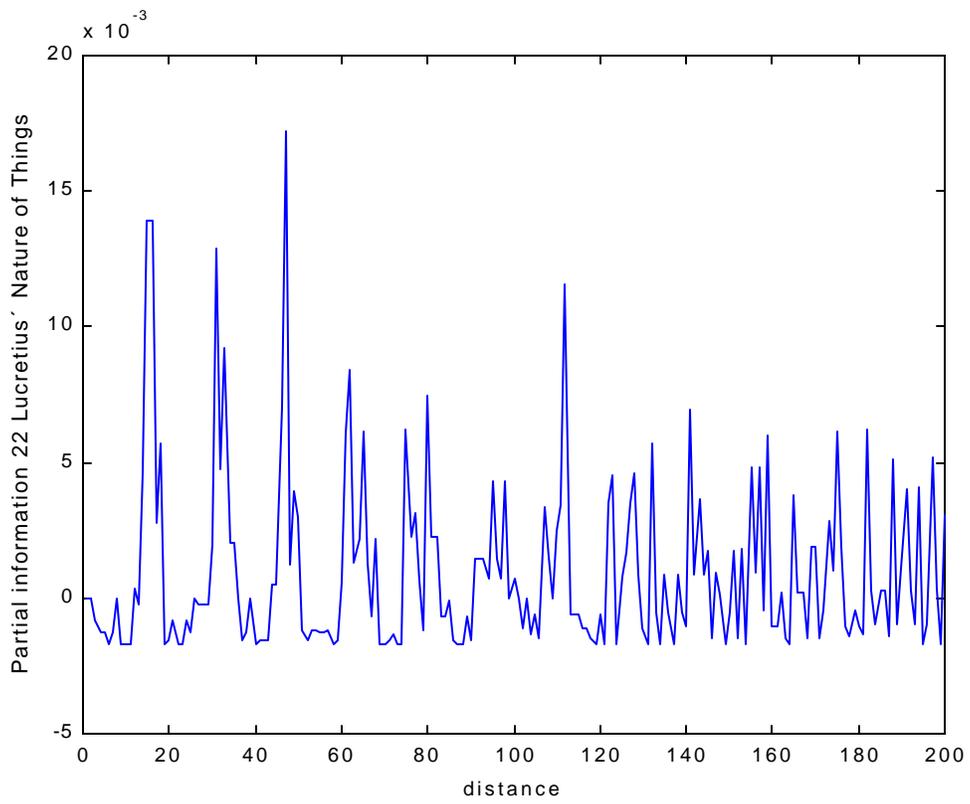

**Fig. 4c Mansilla & Bush**

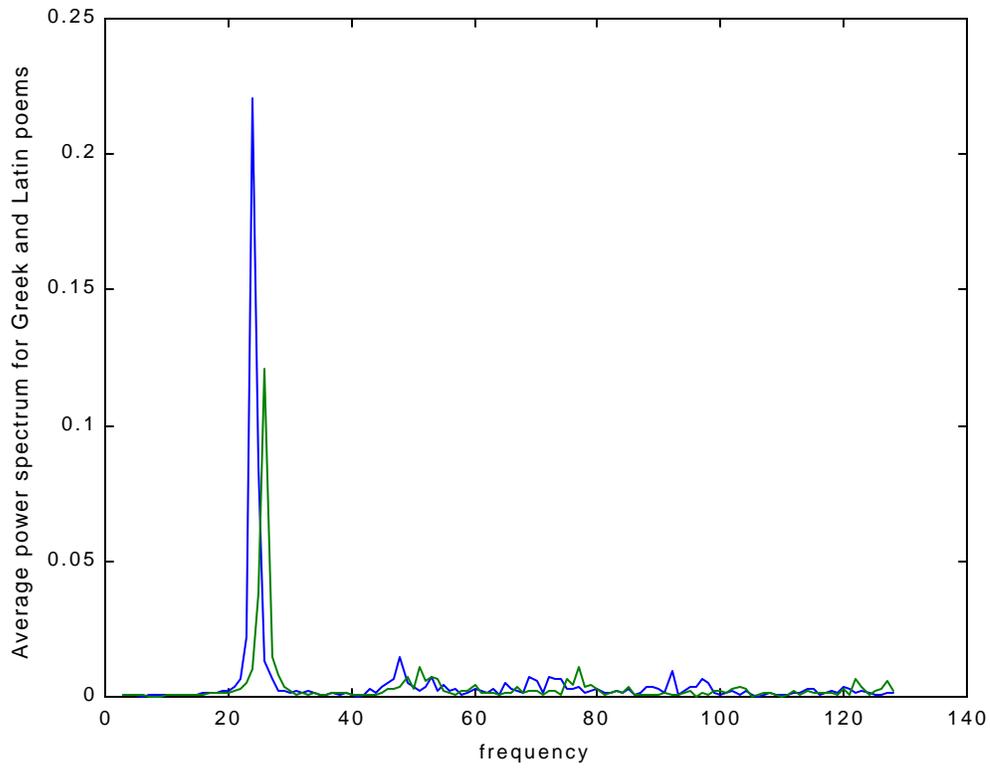

**Fig. 5 Mansilla & Bush**